\documentclass[titlepage,12pt]{article}
\usepackage{amssymb,psfig,epsfig,epsf}

\newcommand{\bea}{\begin{eqnarray}}
\newcommand{\eea}{\end{eqnarray}}
\newcommand{\be}{\begin{equation}}
\newcommand{\ee}{\end{equation}}

\newcommand{\bfsig}{{\mbox{\boldmath $\sigma$}}}

\newcommand{\one}{1\!\mbox{l}}

\begin{document}

\begin{flushright}
Saclay/SPhT-T99/132\\
November 1999\\
\end{flushright}

\vskip 1truecm
 
\begin{center}
{\large\bf POLARIZATION AND SPIN CORRELATIONS  \\[0.4cm]
 OF TOP QUARKS AT A FUTURE e$^+$e$^-$ LINEAR\\[0.4cm]COLLIDER}

\vskip .8truecm

A. Brandenburg$^1$,
M. Flesch$^2$,
P. Uwer$^3$

\end{center}
 
\vskip 1truecm
 
\begin{center}
{\bf ABSTRACT }
\end{center}
We discuss the polarization 
and spin correlations of top quarks produced above threshold
at a future linear collider, including QCD radiative corrections.  
\noindent

\vskip 2.5truecm
\begin{center}
{\it Talk presented by A. Brandenburg at the 
`International Workshop on "Symmetry and Spin",
PRAHA-SPIN99, Prague, Czech Republic, September 5 -12, 1999'}
\end{center}
\vfil

{\footnotesize

\noindent
$^1$ DESY Theory Group, D-22603 Hamburg, Germany.

\noindent
$^2$  Institut f\"ur Theoretische Physik, RWTH Aachen, Germany
D-52056 Aachen.

\noindent
$^3$  Service de Physique Th\'eorique,
Centre d'Etudes de Saclay, F-91191 Gif-sur-Yvette cedex, France.

}

\eject
\setcounter{footnote}{0}
\renewcommand{\thefootnote}{\arabic{footnote}}

\section{Introduction}
Top quark production at a future $e^+e^-$ linear collider
provides an excellent possibility to study
polarization phenomena of quarks without hadronization ambiguities
in a `clean' environment.
Due to its short lifetime, the top quark decays as a quasi-free quark, 
before hadronization effects can take place. The large
width of the top quark thus serves effectively as a cutoff for 
non-perturbative effects.
Information on the polarization and spin correlations of top quarks
is therefore not diluted by hadronization effects but 
transferred to the decay products.
This means that the underlying dynamics of both the production and decay
process of the heaviest elementary particle known to date
can be studied in greater detail, leading to either a confirmation
of the Standard Model (SM) predictions or to hints for `new' physics. 
For example, the chirality structure of the $tWb$ vertex 
can be tested with a highly polarized top quark sample \cite{JeKu94}.
Further, anomalous CP-violating dipole form factors
contributing to the $Z t\bar{t}$ and  $\gamma t\bar{t}$ vertex
would show up as nonzero expectation values of CP-odd spin 
observables (see, e.g., \cite{BeNaOvSc92}).  
Needless to say, the predictions of the SM must be known to high
precision in order to establish possible deviations.\par 
The polarization and spin correlations of top quarks
can be traced in the angular-energy distributions and
momentum correlations of the decay products.
Consider for example the decay distribution of charged
leptons in semileptonic decays 
$t\to \ell^+ \nu_{\ell} b$. At leading order within 
the SM, this distribution reads 
in the top quark rest frame \cite{JeKu89}
\bea\label{decay}
\frac{d^2\Gamma}{dE_{\ell}d\cos\theta}=\frac{1}{2}\left(1+|{\bf P}_t|
\cos\theta\right)\frac{d\Gamma}{dE_{\ell}},
\eea
where $E_{\ell}$ is the energy of the charged lepton and 
$\theta$ is the angle between the direction of $\ell^+$ 
and the polarization ${\bf P}_t$ of the top quark sample.
A remarkable feature of (\ref{decay}) is the factorization into
an energy-dependent and angular-dependent part, which 
is also respected  to a high degree of accuracy by QCD corrections  
\cite{CzJeKu91}. The direction of flight of the charged lepton 
in the top quark rest frame is thus a perfect analyser of the top
quark polarization.  
Analogously, angular correlations between ${\ell^+}$ and  ${\ell^-}$
efficiently probe spin correlations between $t$ and $\bar{t}$.
Of course, momenta of other 
final state particles in semileptonic
decays as well as in hadronic top decays 
can also be used to probe top quark spin effects.
\par
In the remainder of this
paper we discuss the polarization and spin correlations   
in $e^+e^-\to t\bar{t}X$ to order $\alpha_s$ 
within the SM. 
Neglecting so-called non-factorizable contributions it is 
straightforward to combine our results with the known 
decay distributions of polarized top quarks.

\section{Review of  leading order results}\label{sec2}
In this section we write down in a compact form  
leading order results for the top quark
polarization and the spin correlations between $t$ and $\bar{t}$ in the 
reaction 
  \begin{equation}
    \label{process} e^+(p_+,\lambda_+)+e^-(p_-,\lambda_-)\to (\gamma^\ast,Z^\ast)\to t(k_t)+
    \bar t(k_{\bar t}) + X,
  \end{equation}
  where $\lambda_-$ ($\lambda_+$) denotes the longitudinal 
  polarization  of the electron (positron) beam\footnote{
    For a right-handed electron (positron), $\lambda_{\mp}=+1$.}.
Spin effects of top quarks in reaction (\ref{process}) have been analysed
first in ref. \cite{KuReZe86}. A more recent analysis of spin correlations
at leading order can be found in ref. \cite{PaSh96}, where a so-called
`optimal' spin basis is constructed.  
\par
The top quark polarization is defined as two times the expectation value
of the top quark spin operator  ${\bf S}_t$.
The operator  ${\bf S}_t$ acts on the tensor product of
the $t$ and $\bar{t}$  spin spaces and is given by 
${\bf S}_t= \frac{\bfsig}{2}\otimes \one $, where
the first (second) factor in the tensor product 
refers to the $t$ ($\bar{t}$) spin space. (The spin operator of the
top antiquark is defined by ${\bf S}_{\bar t}= \one \otimes \frac{\bfsig}{2}$.)
The expectation value is taken with respect to the spin degrees
of freedom of the $t\bar{t}$ sample described 
by a spin density matrix $\rho$,
i.e.
\bea \label{pol}
{\bf P}_t = 2\,\langle {\bf S}_t\rangle = 
2\frac{{\rm Tr}\, \left[\rho\, {\bf S}_t\right]}{{\rm Tr}\, \rho}.
\eea  
For details on the definition and computation of $\rho$, see \cite{BrFlUw99}.
The polarization of the top antiquark ${\bf P}_{\bar t}$ 
is defined by replacing ${\bf S}_t$ by ${\bf S}_{\bar t}$ in (\ref{pol}).
For top quark pairs produced by CP invariant interactions, 
${\bf P}_{\bar t}={\bf P}_{t}$.
The spin correlations between
$t$ and $\bar{t}$ are encoded in the matrix
\bea \label{corr} 
C_{ij} = 4\,\langle S_{t,i} S_{\bar{t},j} \rangle = 
4\frac{{\rm Tr}\,\left[ \rho \,S_{t,i}S_{\bar{t},j}\right]}{{\rm Tr}\, \rho}.
\eea
The definitions (\ref{pol}) and (\ref{corr})
imply that ${\bf P}_t$ and $C_{ij}$ are independent
of the choice of the spin basis. 
It is convenient to write the results in terms of the electron and top quark
 directions
$\hat{\bf p}$ and $\hat{\bf k}$ defined in the c.m. system, the
cosine of the scattering angle $z=\hat{\bf p}\cdot\hat{\bf k}$, the 
scaled top quark mass $r=2m_t/\sqrt{s}$ and the top quark velocity
$\beta=\sqrt{1-r^2}$. 
The electroweak couplings that enter
the results are given by 
  \begin{eqnarray}
    \label{wcouplings}
    g_{PC (PV)}^{VV} &=&Q_t^2\, f_{PC(PV)}^{\gamma\gamma} 
    + 2 \,g_v^t\,Q_t\, \chi\,f_{PC(PV)}^{\gamma Z} 
    + g_v^{t\,2}\, \chi^2\,f_{PC(PV)}^{ZZ},\nonumber\\
    g_{PC(PV)}^{AA}  &=& g_a^{t\,2} \chi^2 f_{PC(PV)}^{ZZ},
    \nonumber\\
    g_{PC(PV)}^{VA}&=& -g_a^t\,Q_t\,\chi\,f_{PC(PV)}^
    {\gamma Z} -g_v^t\,g_a^t\, \chi^2 f_{PC(PV)}^{ZZ},
  \end{eqnarray}
  where
  \bea
      f_{PC}^{\gamma\gamma}&=&1-\lambda_-\lambda_+,\nonumber \\
      f_{PV}^{\gamma\gamma}&=& \lambda_--\lambda_+,\nonumber \\
      f_{PC}^{ZZ}&=&(1-\lambda_-\lambda_+)(g_v^{e2}+g_a^{e2})-
      2(\lambda_--\lambda_+) g_v^{e} g_a^{e},\nonumber \\
      f_{PV}^{ZZ}&=&(\lambda_--\lambda_+) (g_v^{e\,2}+g_a^{e\,2}) -
      2\,(1-\lambda_-\lambda_+)g_v^e\,g_a^e,\nonumber \\
      f_{PC}^{\gamma Z}&=&-(1-\lambda_-\lambda_+)g_v^e + 
      (\lambda_--\lambda_+)g_a^e,\nonumber \\
      f_{PV}^{\gamma Z}&=& (1-\lambda_-\lambda_+)g_a^e -(\lambda_--\lambda_+)
      g_v^e.
  \eea 
In (\ref{wcouplings}),
 $Q_t$ denotes 
  the electric charge of the top quark in units of $e=\sqrt{4\pi\alpha}$, and
  $g_v^f$, $g_a^f$ are the vector- and the axial-vector couplings of a
  fermion of type $f$, i.e.
  $g_v^e = -\frac{1}{2} + 2 \sin^2\vartheta_W$, 
  $g_a^e =-\frac{1}{2}$ for an electron, and  
  $g_v^t = \frac{1}{2} - \frac{4}{3} \sin^2\vartheta_W$,
  $g_a^t = \frac{1}{2}$ for a top quark, with $\vartheta_W$ denoting the 
  weak mixing angle. The function
  $\chi$ is given by
  \begin{equation}
    \label{chi}
    \chi = \frac{1}{4\sin^2\vartheta_W\cos^2\vartheta_W}\,
    \frac{s}{s-m_Z^2},
  \end{equation}
  where $m_Z$ stands for the mass of the Z boson.
\par
We further introduce a vector perpendicular to ${\bf k}$ in the
production plane,
${\bf k}^{\perp}=\hat{\bf p}-z\hat{\bf k}$. A simple calculation yields: 
\bea
{\bf P}_t
&\!=&\!
2\frac{ r\left(\beta z  g_{PC}^{VA}+g_{PV}^{VV}\right){\bf k}^{\perp}+\left[\beta (1+z^2 )g_{PC}^{VA}
+z  g_{PV}^{VV}+\beta^2 z  g_{PV}^{AA}\right] \hat{\bf k} }
{ \left[2-\beta^2 (1-z^2)\right]g_{PC}^{VV}+\beta^2 
(1+z^2 )g_{PC}^{AA}+4\beta z g_{PV}^{VA} },
\eea
\newpage
\bea
C_{ij}
&\!=&\!  \frac{1}{3}\delta_{ij}
+\frac{2}{\left[2-\beta^2 (1-z^2)\right]
g_{PC}^{VV} +\beta^2 (1+z^2) g_{PC}^{AA}
+4\beta z g_{PV}^{VA}} \nonumber \\
&\!\times&\!
\bigg[ \left(\left[z^2 +\beta^2(1-z^2)\right]g_{PC}^{VV}+
\beta^2 z^2 g_{PC}^{AA}+2\beta z g_{PV}^{VA} \right) 
\left(\hat{k}_i\hat{k}_j-\frac{1}{3}\delta_{ij}\right)
\nonumber \\ & & +\,
(g_{PC}^{VV}-\beta^2 g_{PC}^{AA}) 
\left(k^{\perp}_i k^{\perp}_j-\frac{1}{3}\delta_{ij}
(1-z^2)\right)
\nonumber \\ 
& & +\, 
r (z g_{PC}^{VV}+ \beta g_{PV}^{VA})
(k^{\perp}_i\hat{k}_j+k^{\perp}_j\hat{k}_i)\bigg].
\eea
In the limit $\beta\to 0$ 
(threshold)  we obtain for the top quark polarization:
\bea \label{plimit}
{\bf P}_t & \!{\buildrel
\beta\to 0\over \longrightarrow}&\! \frac{g_{PV}^{VV}}{g_{PC}^{VV}}\hat{\bf p}
+  \beta\left[\left(
\frac{g_{PC}^{VA}}{g_{PC}^{VV}}-2\,\frac{g_{PV}^{VV}g_{PV}^{VA}}{(g_{PC}^{VV})^2}\right)z\,\hat{\bf p} + \frac{g_{PC}^{VA}}{g_{PC}^{VV}}\hat{\bf k} \right] + 
{\cal O}(\beta^2).
\eea
In the leading order 
parton model calculation, the 
top quark polarization becomes parallel to the electron
beam for $\beta=0$. 
For a fully polarized electron beam
(and unpolarized positrons), we
have $g_{PV}^{VV(VA)}=\pm g_{PC}^{VV(VA)}$ for $\lambda_-=\pm 1$. 
In that case the top quark polarization along the beam 
is equal to the electron polarisation, 
${\bf P_t}\cdot \hat{\bf p}=\lambda_-=\pm 1$, 
up to corrections of order $\beta^2$.
\par
The spin correlations also have a simple limit:
\bea \label{corrlimit}
C_{ij}
&\!{\buildrel
\beta\to 0\over \longrightarrow}&\! \hat{p}_i\hat{p}_j+
\beta \frac{g_{PV}^{VA}}{g_{PC}^{VV}}(\hat{p}_i  \hat{k}_j + 
 \hat{p}_j  \hat{k}_i- 2 z  \hat{p}_i  \hat{p}_j)
+{\cal O}(\beta^2).
\eea 
Note that in the threshold region
QCD binding effects modify the above parton model 
results significantly. More precisely,
the simple factor $\beta$ in (\ref{plimit}) and (\ref{corrlimit}) 
gets replaced by a function incorporating the complex dynamics of the 
$t\bar{t}$ system close to threshold, which is  
governed by the QCD potential \cite{Ha95}. 
\par
In the high-energy limit $r=\sqrt{1-\beta^2}\to 0$,
\bea
{\bf P}_t
{\buildrel
r\to 0\over \longrightarrow} \,2\,\frac{
(1+z^2 )g_{PC}^{VA}+ z  (g_{PV}^{VV}+g_{PV}^{AA})}{(1+z^2 )
(g_{PC}^{VV}+g_{PC}^{AA})+4 z g_{PV}^{VA} } \hat{\bf k} + {\cal O}(r),
\eea
i.e. the top quark polarization becomes parallel to its direction of flight. 
\newpage
Finally,
\bea
C_{ij}
 &\!{\buildrel
r\to 0\over \longrightarrow}&\!
\frac{1}{3}\delta_{ij} + \frac{2}
{(1+z^2)(g_{PC}^{VV}+g_{PC}^{AA})+4 g_{PV}^{VA} z }\nonumber \\
&\!\times&\!
\bigg[(g_{PC}^{VV}+z^2  g_{PC}^{AA}+2 z g_{PV}^{VA} )
\left(\hat{k}_i\hat{k}_j-\frac{1}{3}\delta_{ij}\right) \nonumber \\
& & +\, (g_{PC}^{VV}-g_{PC}^{AA} )\left(k^{\perp}_i 
k^{\perp}_j-\frac{1}{3}\delta_{ij}
(1-z^2)\right)\bigg] + {\cal O}(r).
\eea
\section{QCD corrections at order $\alpha_s$}
The QCD corrections at order $\alpha_s$ 
to the above results are given by the contributions from
one-loop virtual corrections to $e^+e^-\to t\bar{t}$ 
and from the real gluon emission process  $e^+e^-\to t\bar{t}g$.
The so-called phase space slicing method is used to isolate
the soft gluon singularities. 
The contribution of hard gluons to ${\bf P}_t$ and $C_{ij}$
is computed by numerically integrating
all phase space variables of the  $t\bar{t}g$ final state 
except for the top quark scattering angle.  
Further details of the computation are given
in ref. \cite{BrFlUw99}.
Results to order $\alpha_s$ for the 
polarization projected onto $\hat{\bf k}$ and 
${\bf k}^{\perp}/|{\bf k}^{\perp}|$
can also be found in  ref. \cite{KoPiTu94} and  
ref. \cite{GrKo96}, respectively.

Absorptive parts of the one-loop amplitude induce as new structures
a polarization normal to the event plane \cite{KuReZe86,KaPuRe78,BeMaSc92} 
as well as new types
of spin correlations. We denote these additional structures
by an upper index `abs'. 
Defining ${\bf n}= \hat{\bf p}\times \hat{\bf k}$, they read:
\bea 
{\bf P}_t^{\rm abs.}&\! = &\!\frac{\alpha_s C_F r 
\left[(\beta^2-2) g_{PV}^{VA}-\beta z g_{PC}^{VV}\right]}
{2\left(g_{PC}^{VV}\left[2-\beta^2(1-z^2)\right]+g_{PC}^{AA}\beta^2 
(1+z^2)+4  g_{PV}^{VA} \beta z\right)}\, {\bf n} \nonumber \\
&\!{\buildrel
\beta\to 0\over \longrightarrow}&\! = 
-\frac{\alpha_s C_F}{2} \frac{g_{PV}^{VA}}{g_{PC}^{VV}}
\,{\bf n}+{\cal O}(\beta),
\eea

\bea
C_{ij}^{\rm abs.}
&\!=&\! \frac{-\alpha_s C_F r }{2\left(g_{PC}^{VV}\left[2-\beta^2(1-z^2)\right]
+g_{PC}^{AA}\beta^2 
(1+z^2)+4  g_{PV}^{VA} \beta z\right)}\nonumber \\
&\!\times&\! \big[(\beta g_{PV}^{VV}-(\beta^2 - 2) z g_{PC}^{VA})
({n}_i\hat{k}_j+\hat{k}_i{n}_j)\nonumber \\
& & +\,
2 r g_{PC}^{VA}
({n}_i k^{\perp}_j+k^{\perp}_i{n}_j)\big]\nonumber \\
 &\!{\buildrel
\beta\to 0\over \longrightarrow}&\! = -\frac{\alpha_s C_F}{2}
\frac{g_{PC}^{VA}}{g_{PC}^{VV}}
\left({n}_i\hat{p}_j+\hat{p}_i{n}_j\right)
+{\cal O}(\beta),
\eea
where $C_F=4/3$.
The threshold behaviour of the other order $\alpha_s$ corrections 
to the Born results for ${\bf P}_t$ and 
$C_{ij}$ is very simple:
First, all these corrections vanish at $\beta=0$.
Second, the QCD corrections of order $\beta$ can be implemented in the Born
formulas (\ref{plimit}) and  (\ref{corrlimit}) by 
multiplying the respective order $\beta$ term with 
the factor $(1+\alpha_s C_F/\pi)$.
\par
%
%
\begin{figure}
\centering
\mbox{\epsfysize=41mm\epsffile{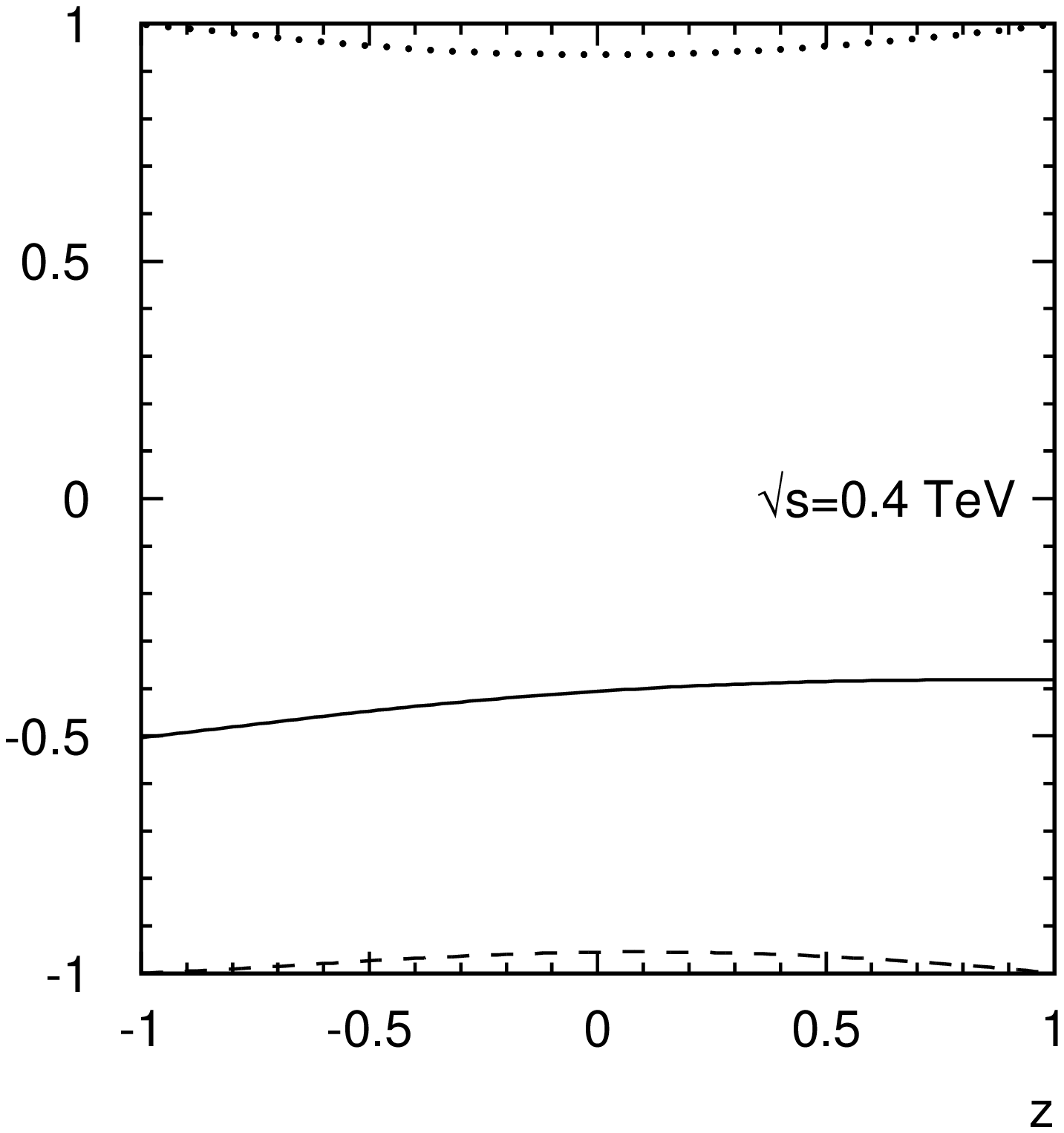}}
\mbox{\epsfysize=41mm\epsffile{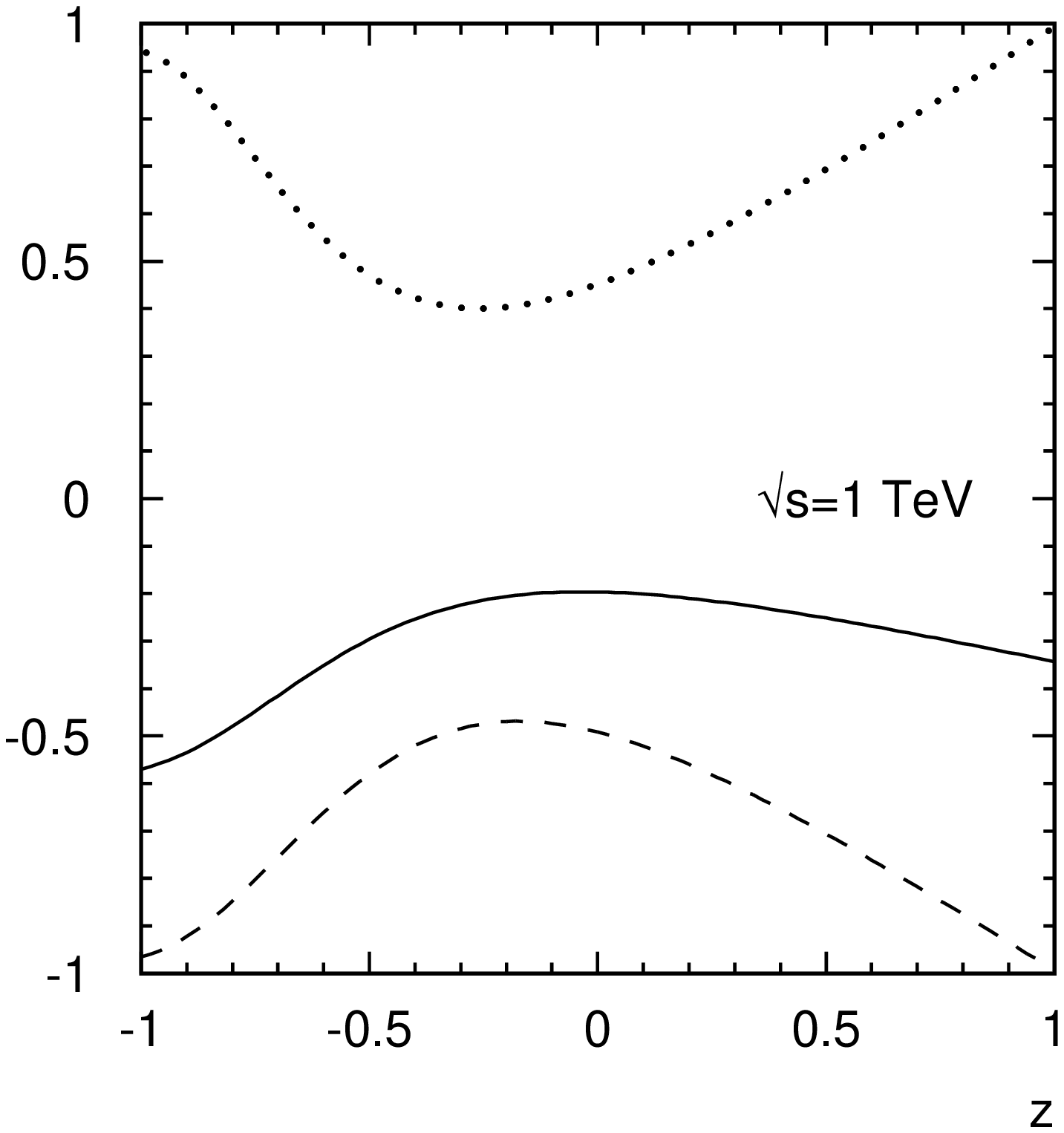}}
\mbox{\epsfysize=41mm\epsffile{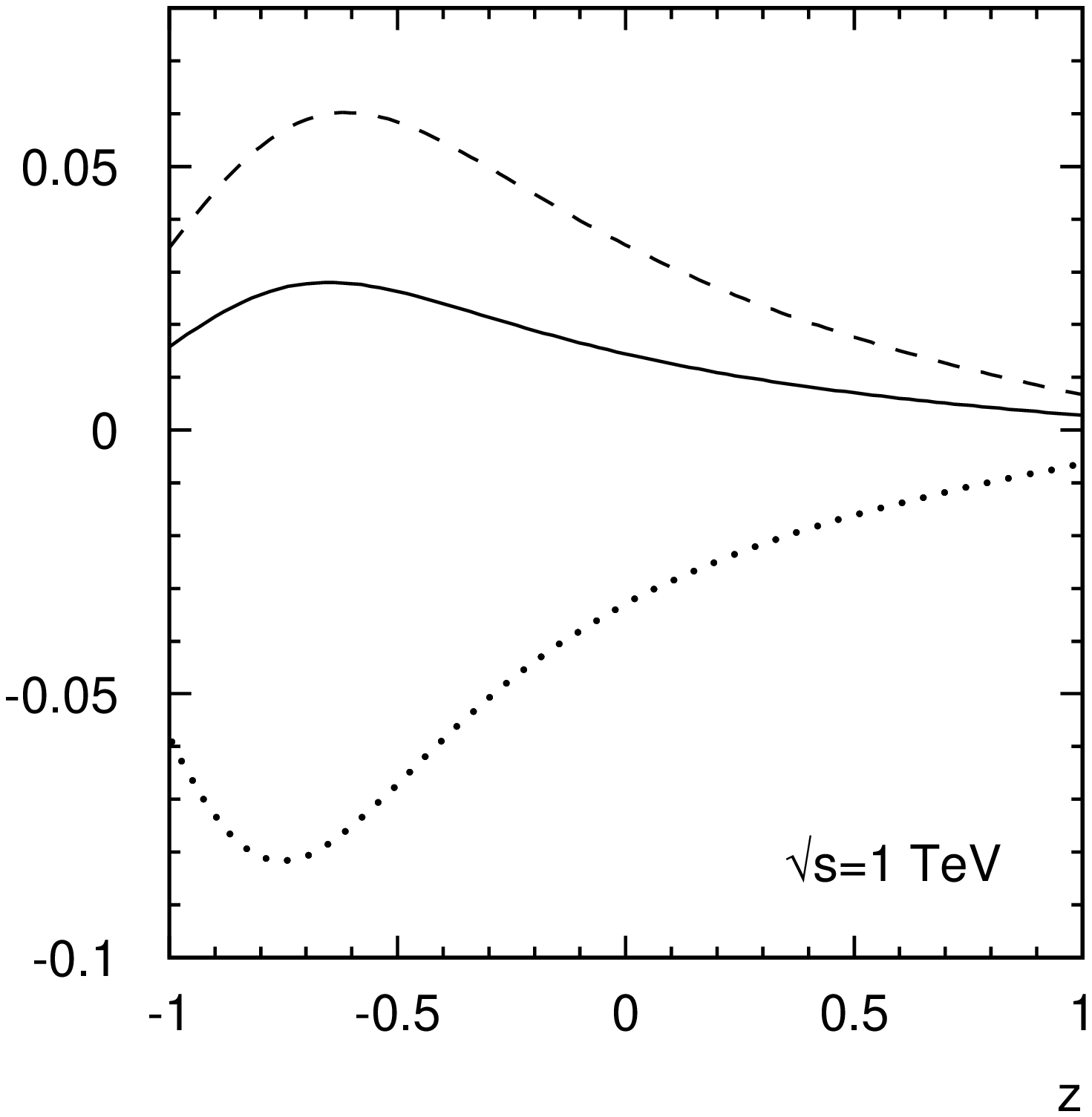}}
\caption{Top quark polarization
projected onto the direction of the electron, i.e. ${\bf P}_t\cdot
\hat{\bf p}$, as a function of $z$. 
The left and middle figure are the results including the order $\alpha_s$
corrections, the 
right figure shows the value of the QCD correction itself 
at $\sqrt{s}=1$ TeV. 
Input values: $m_t=175$ GeV, 
$\alpha_s=0.1$ (fixed), $\sin^2\vartheta_W=0.2236$, and $\lambda_+=0$. 
The solid line is for $\lambda_-=0$, the dashed line for 
$\lambda_-=-1$, and the dotted line for $\lambda_-=+1$.} 
\label{F1}
\end{figure}
%
%
\begin{figure}
\centering
\mbox{\epsfysize=41mm\epsffile{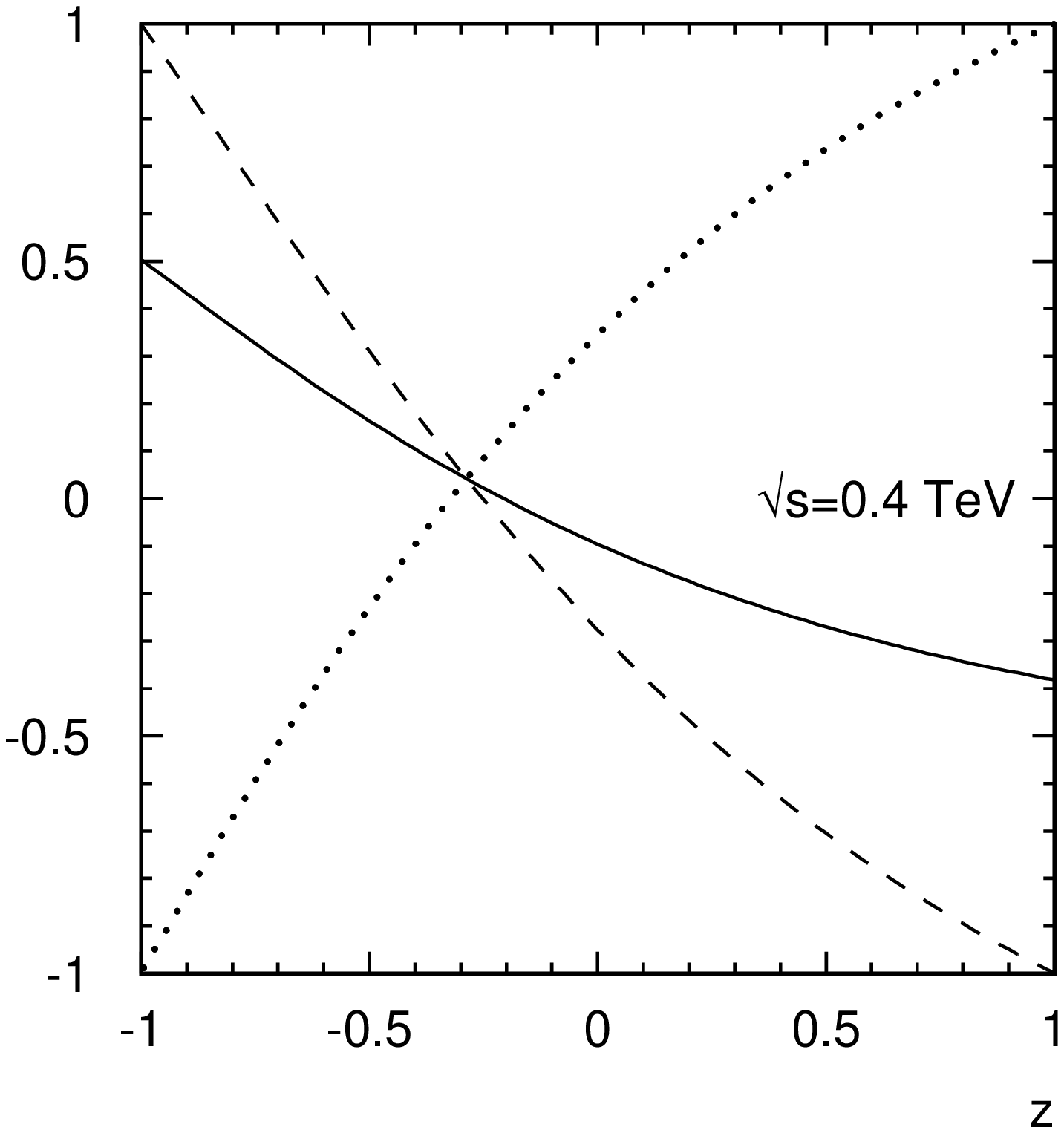}}
\mbox{\epsfysize=41mm\epsffile{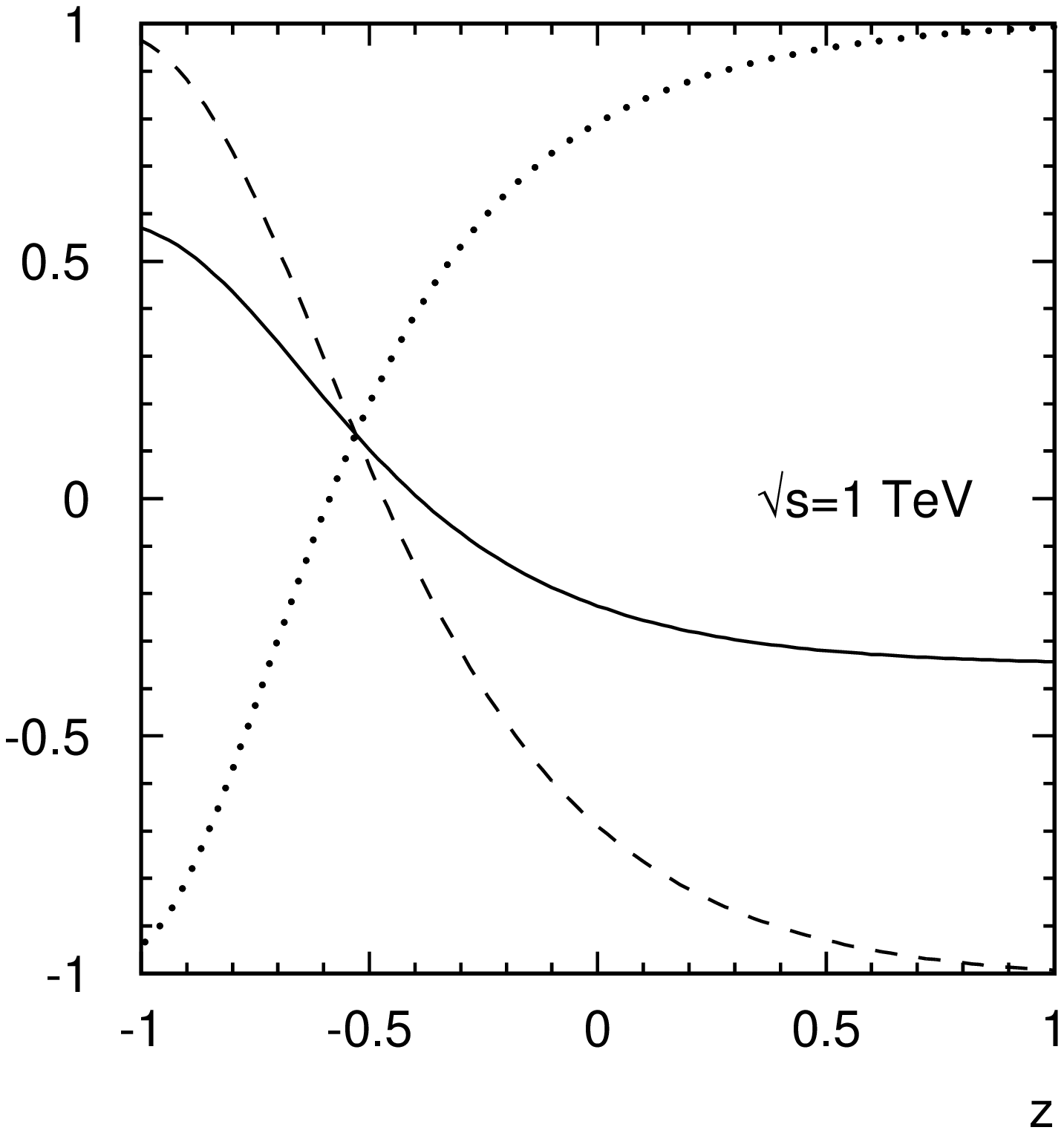}}
\mbox{\epsfysize=41mm\epsffile{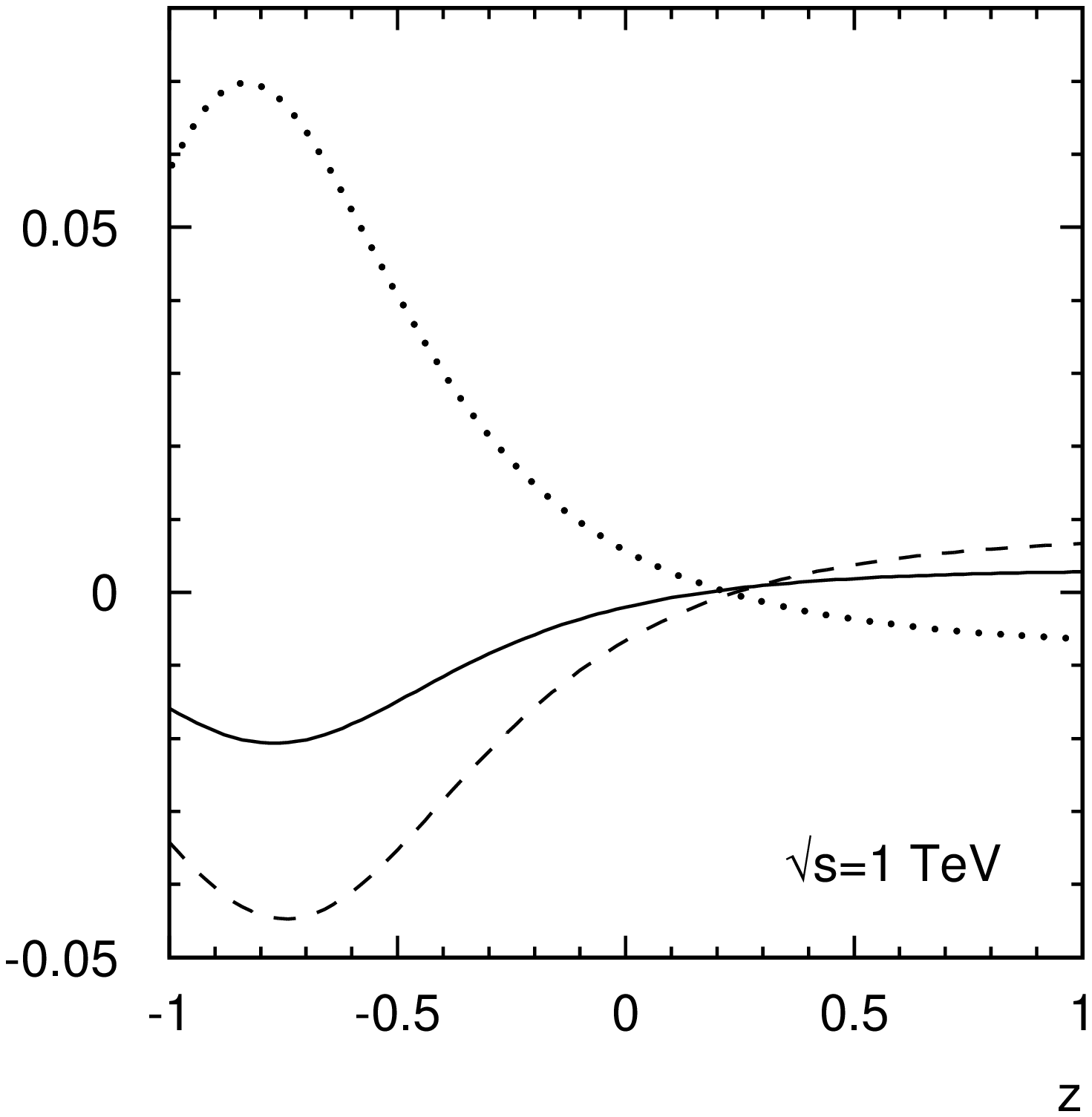}}
\caption{Same as Fig. 1, but for ${\bf P}_t\cdot
\hat{\bf k}$.}
\label{F2}
\end{figure}

%
%
\begin{figure}
\centering
\mbox{\epsfysize=41mm\epsffile{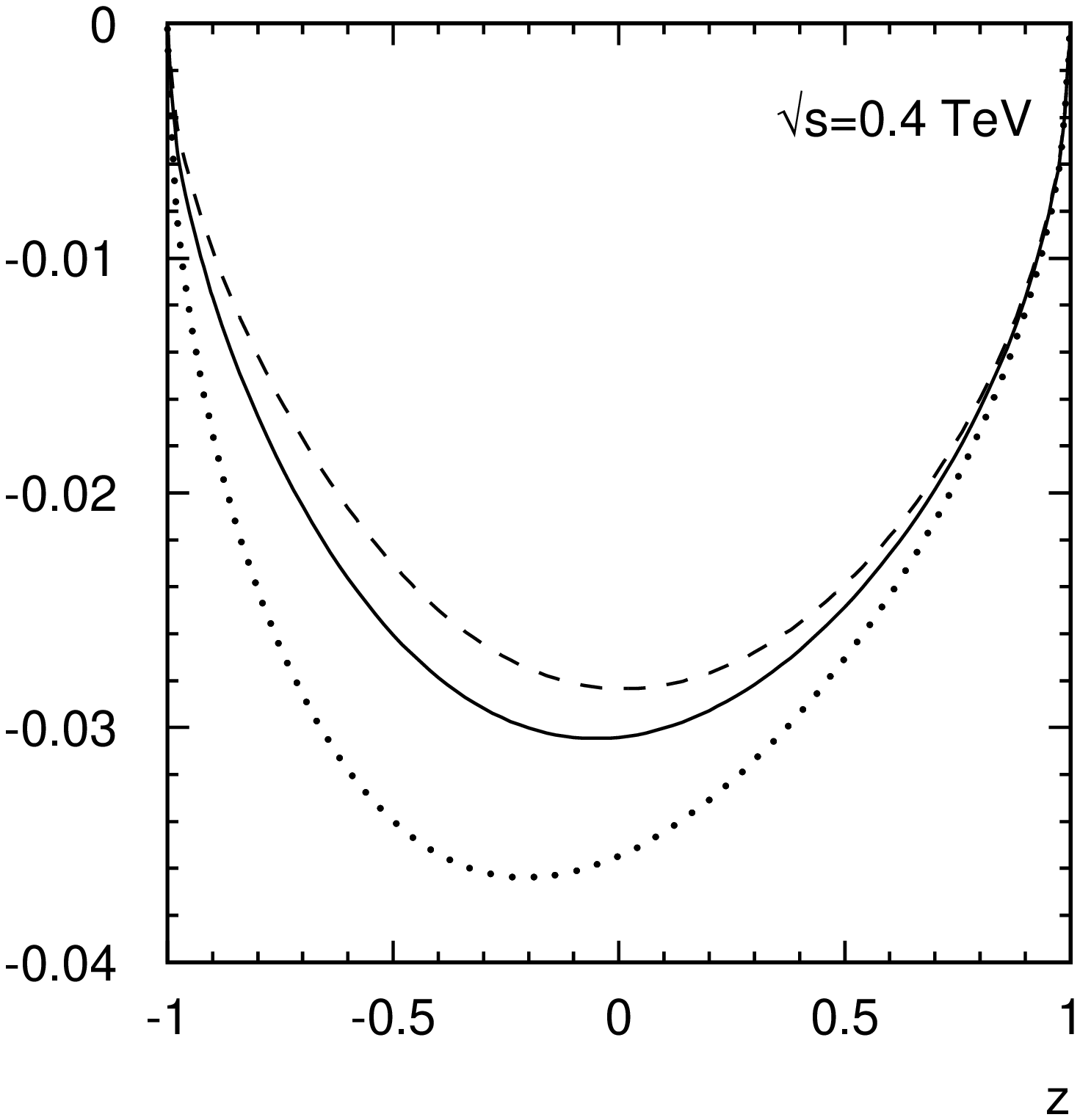}}
\mbox{\epsfysize=41mm\epsffile{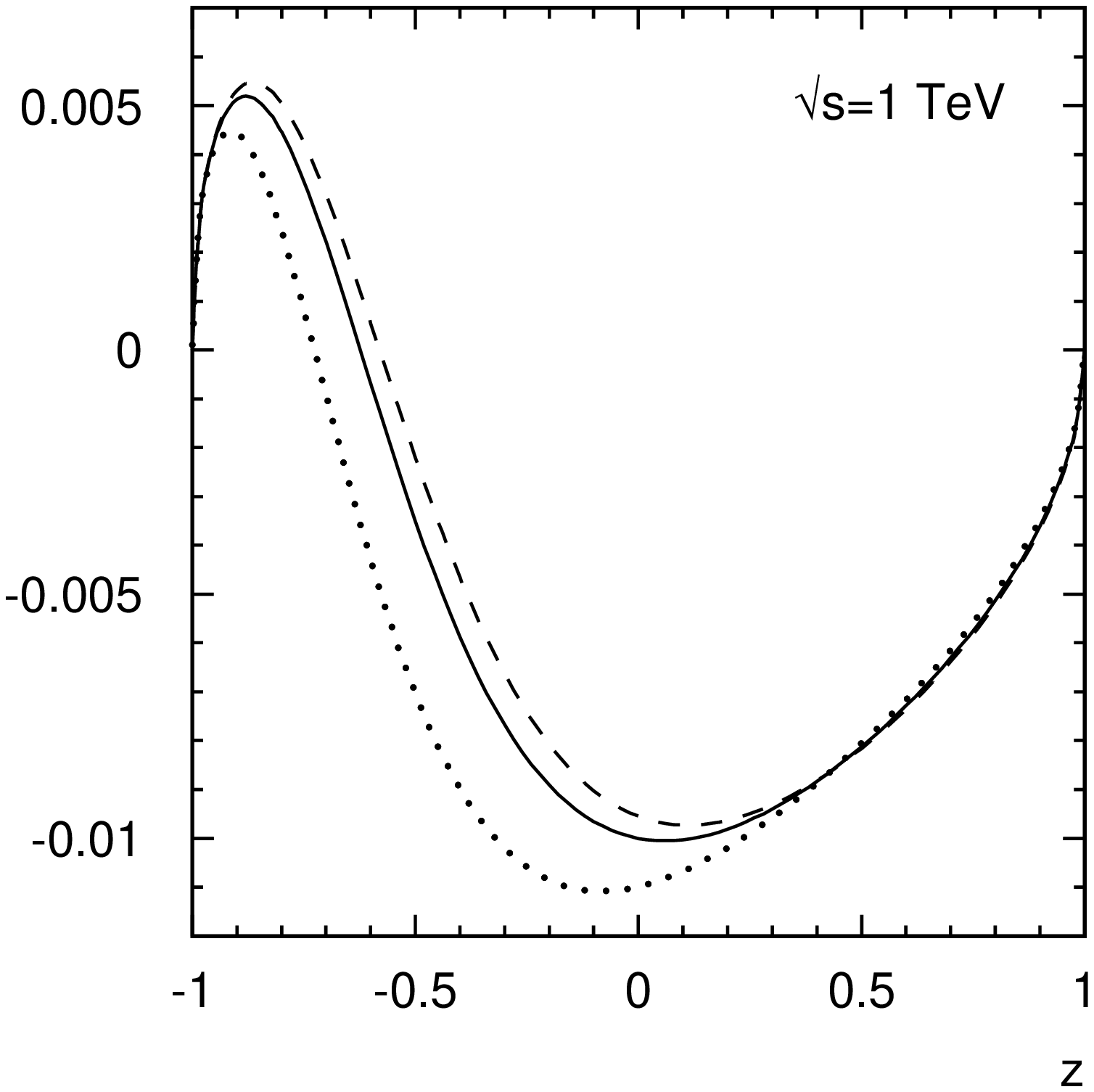}}
\caption{Top quark polarization
projected onto $\hat{\bf n}={\bf n}/|{\bf n}|$, i.e. 
${\bf P}_t\cdot\hat{\bf n}$,  which is zero at Born level.
The labelling of the curves is as in Fig. 1.}
\label{F3}
\end{figure}

%
%
\begin{figure}
\centering
\mbox{\epsfysize=41mm\epsffile{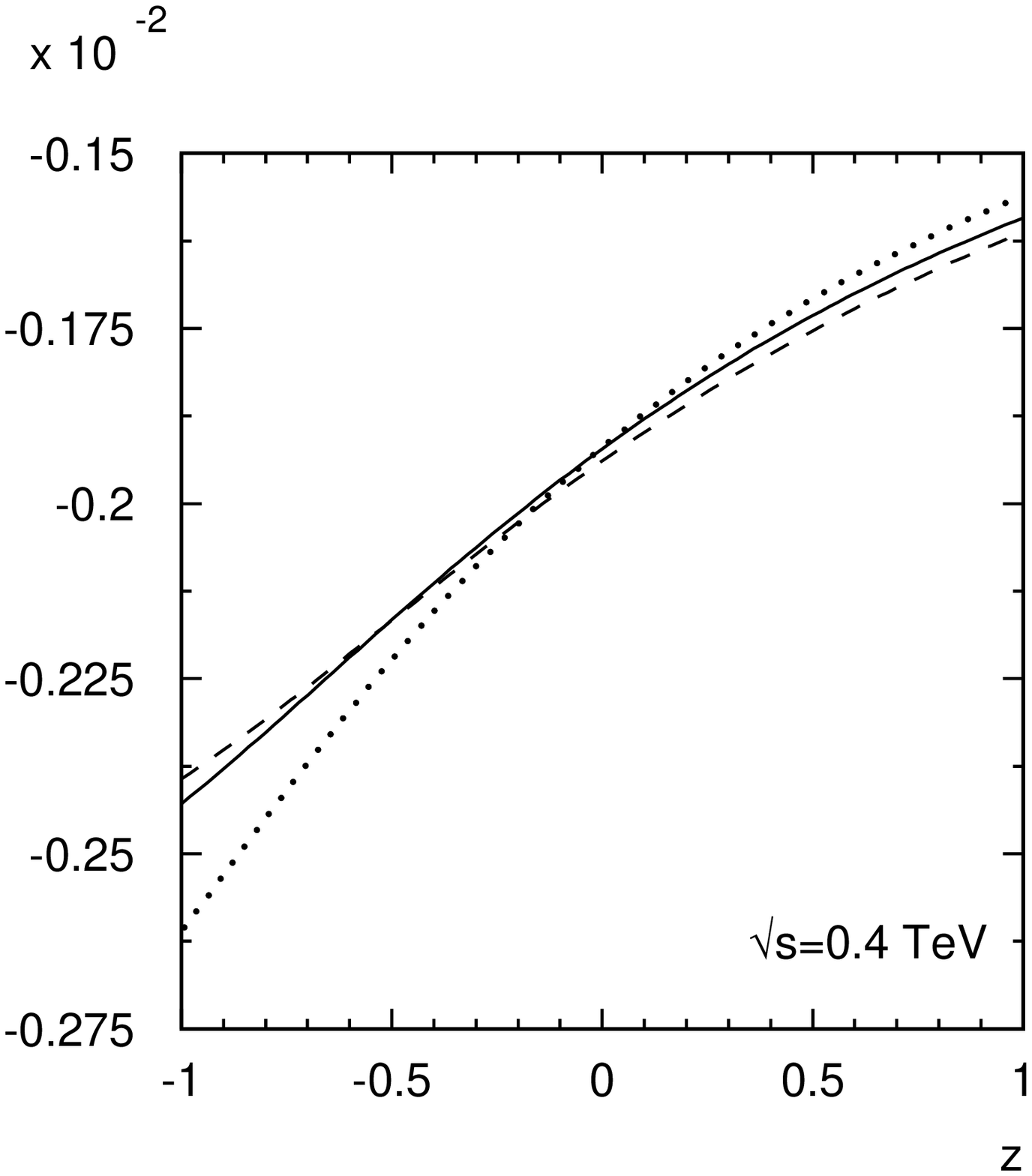}}
\mbox{\epsfysize=41mm\epsffile{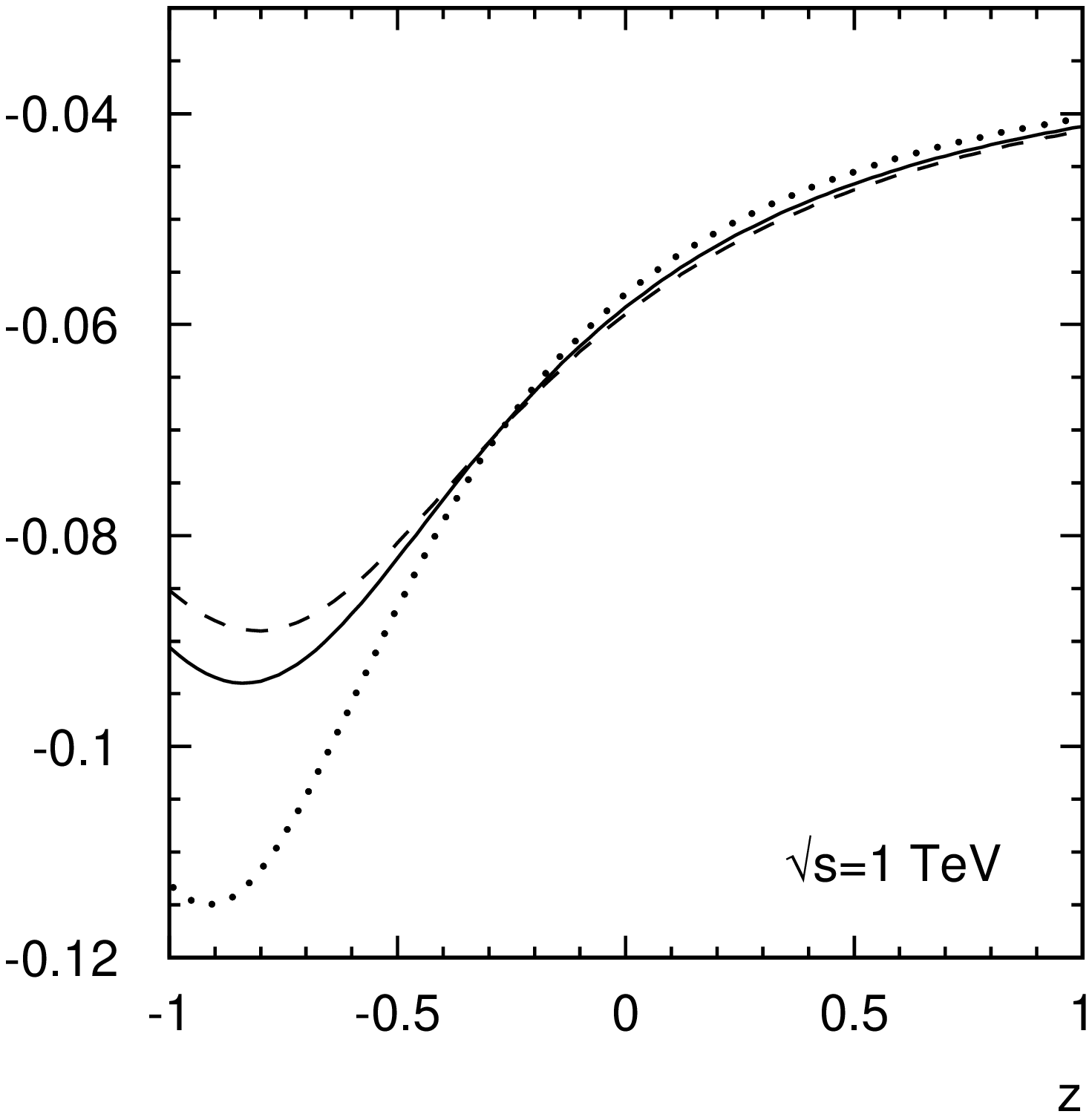}}
\caption{Order $\alpha_s$ correction to the correlation 
$C_{ij}\delta_{ij}=4\langle {\bf S}_t\cdot {\bf S}_{\bar{t}}\rangle$,
which is equal to 1 at the Born level. The labelling of 
the curves is as in Fig. 1.  
}
\label{F4}
\end{figure}
%
%
\begin{figure}
\centering
\mbox{\epsfysize=41mm\epsffile{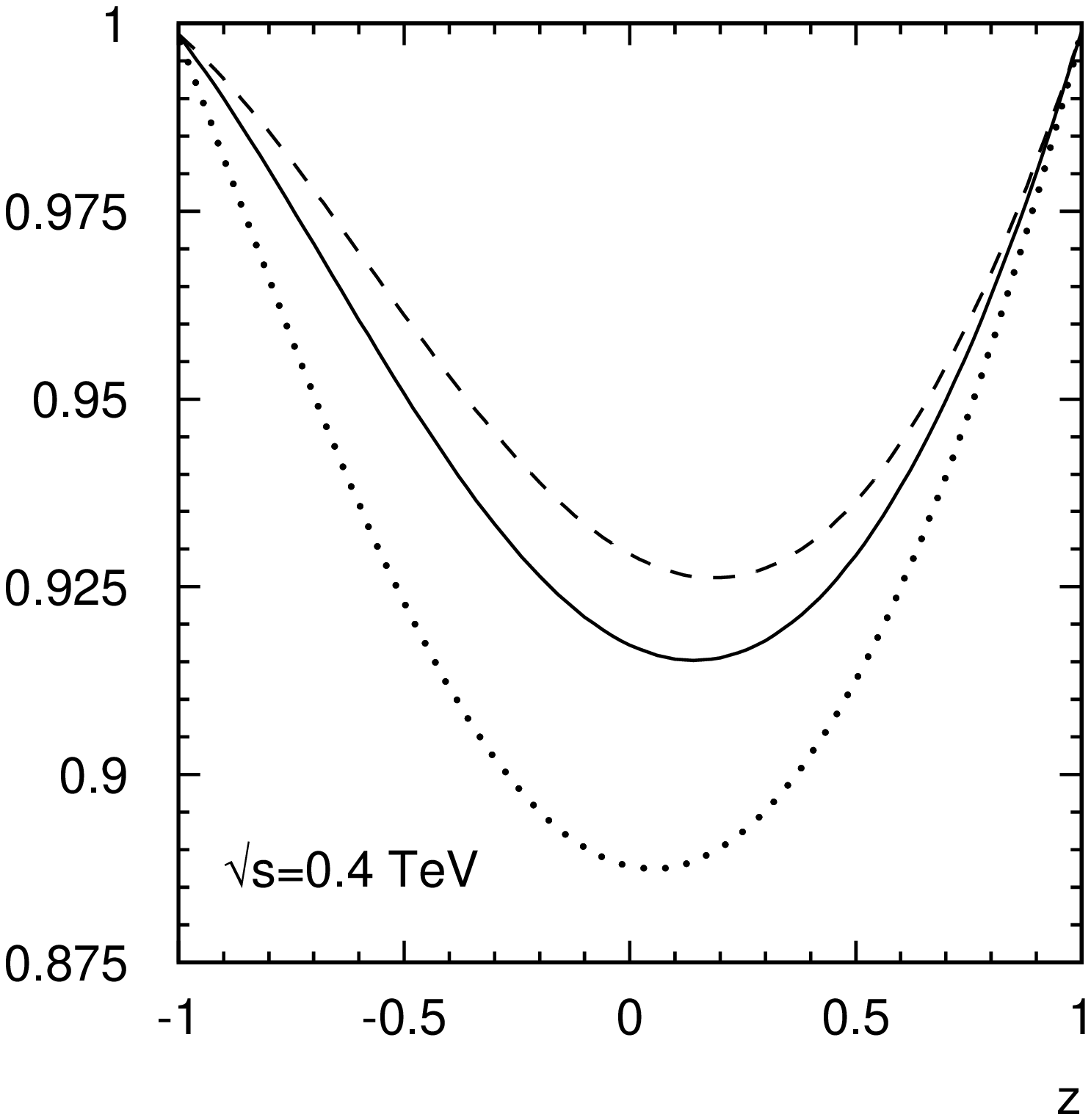}}
\mbox{\epsfysize=41mm\epsffile{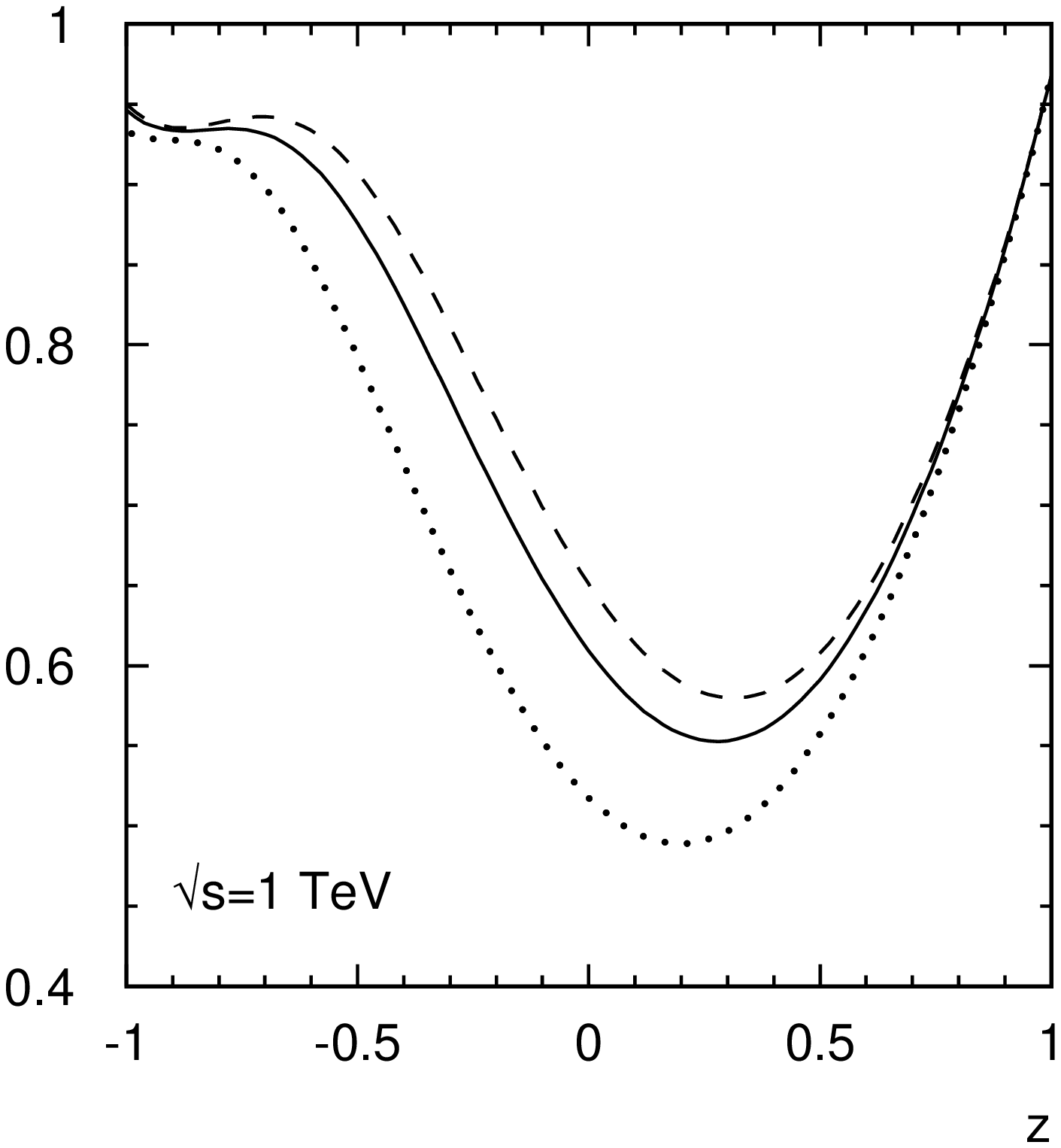}}
\mbox{\epsfysize=41mm\epsffile{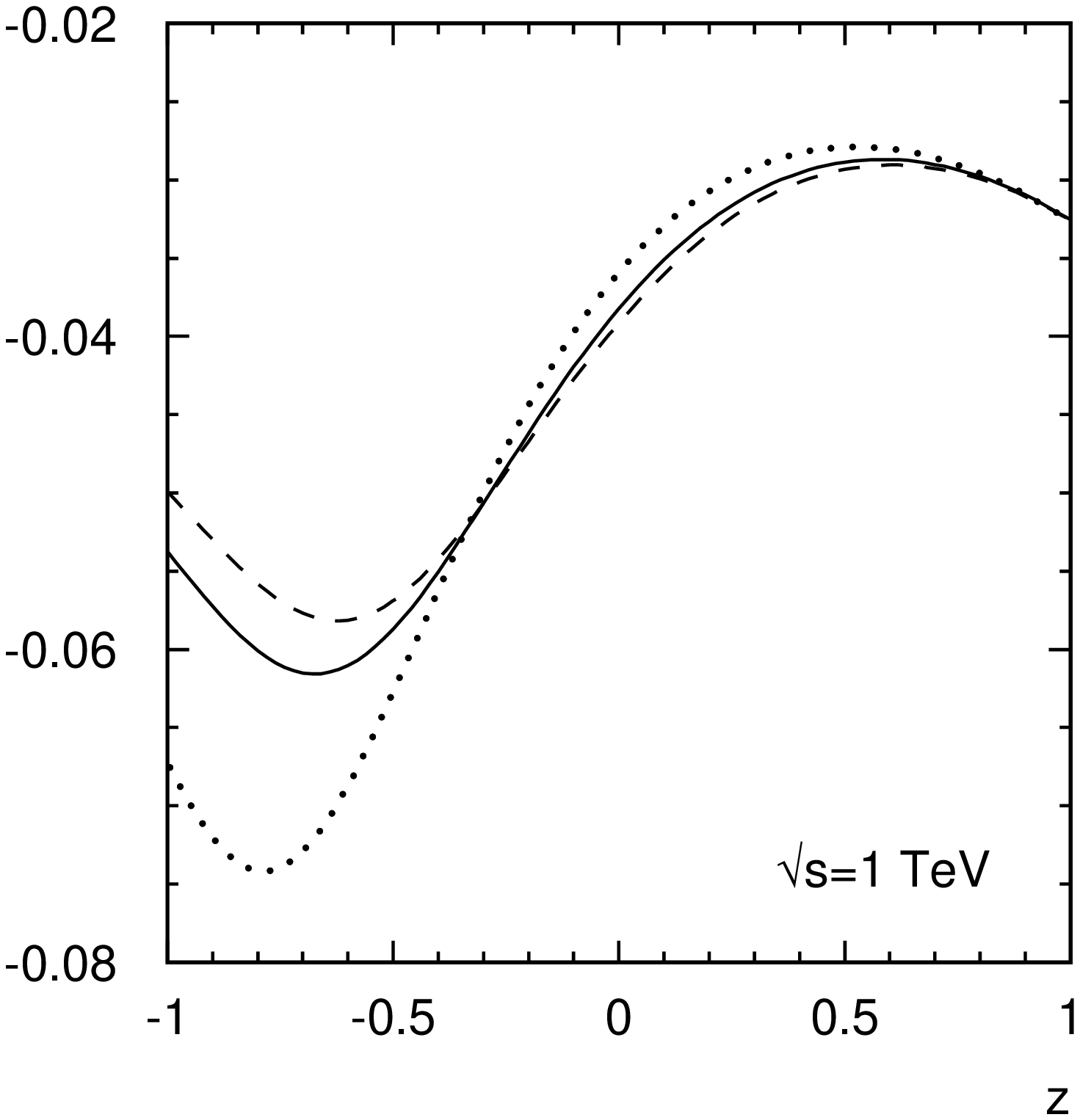}}
\caption{Same as Fig. 1 , but for the correlation
$\hat{p}_iC_{ij}\hat{p}_j$.
}
\label{F5}
\end{figure}

%
%
\begin{figure}
\centering
\mbox{\epsfysize=41mm\epsffile{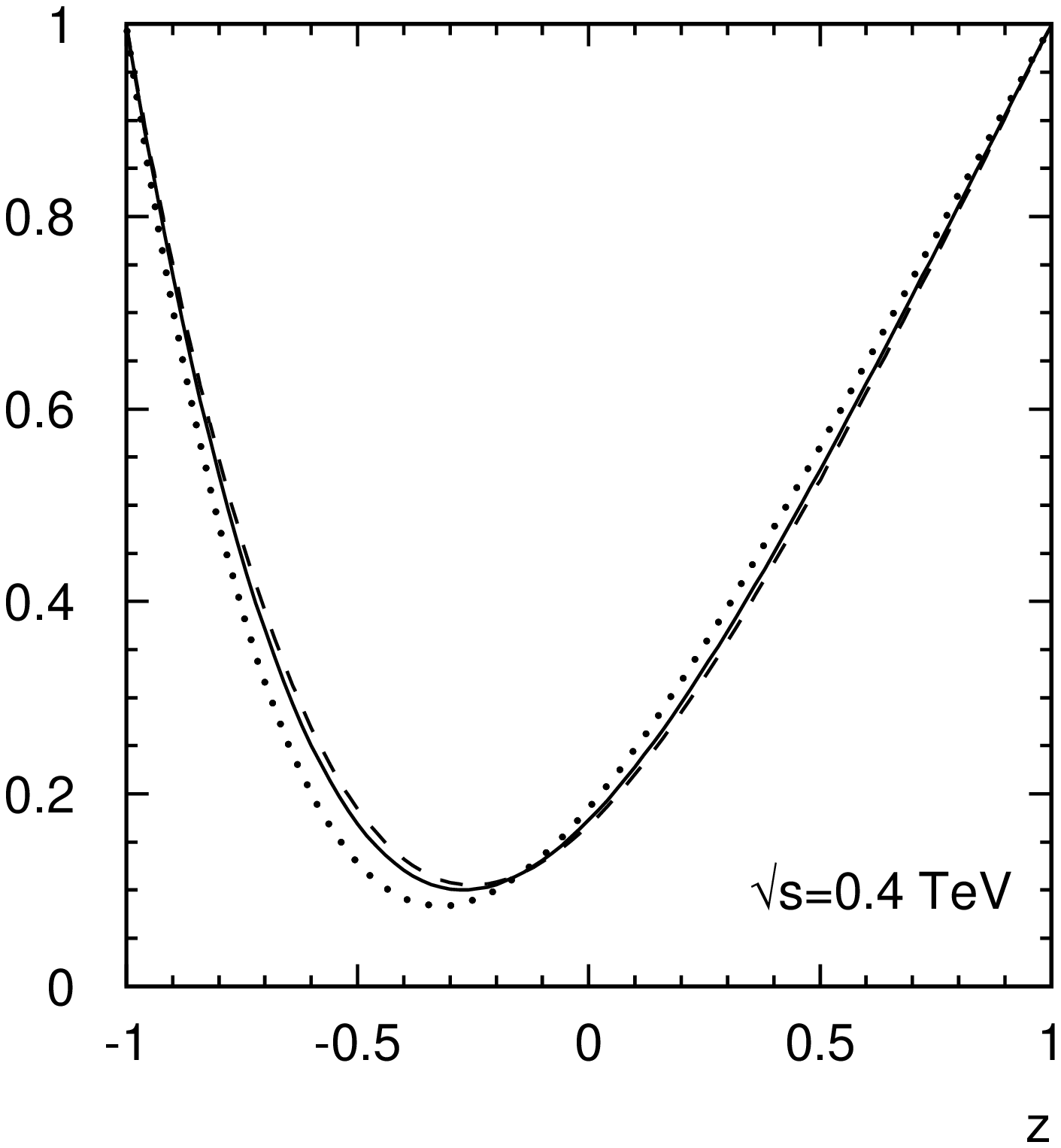}}
\mbox{\epsfysize=41mm\epsffile{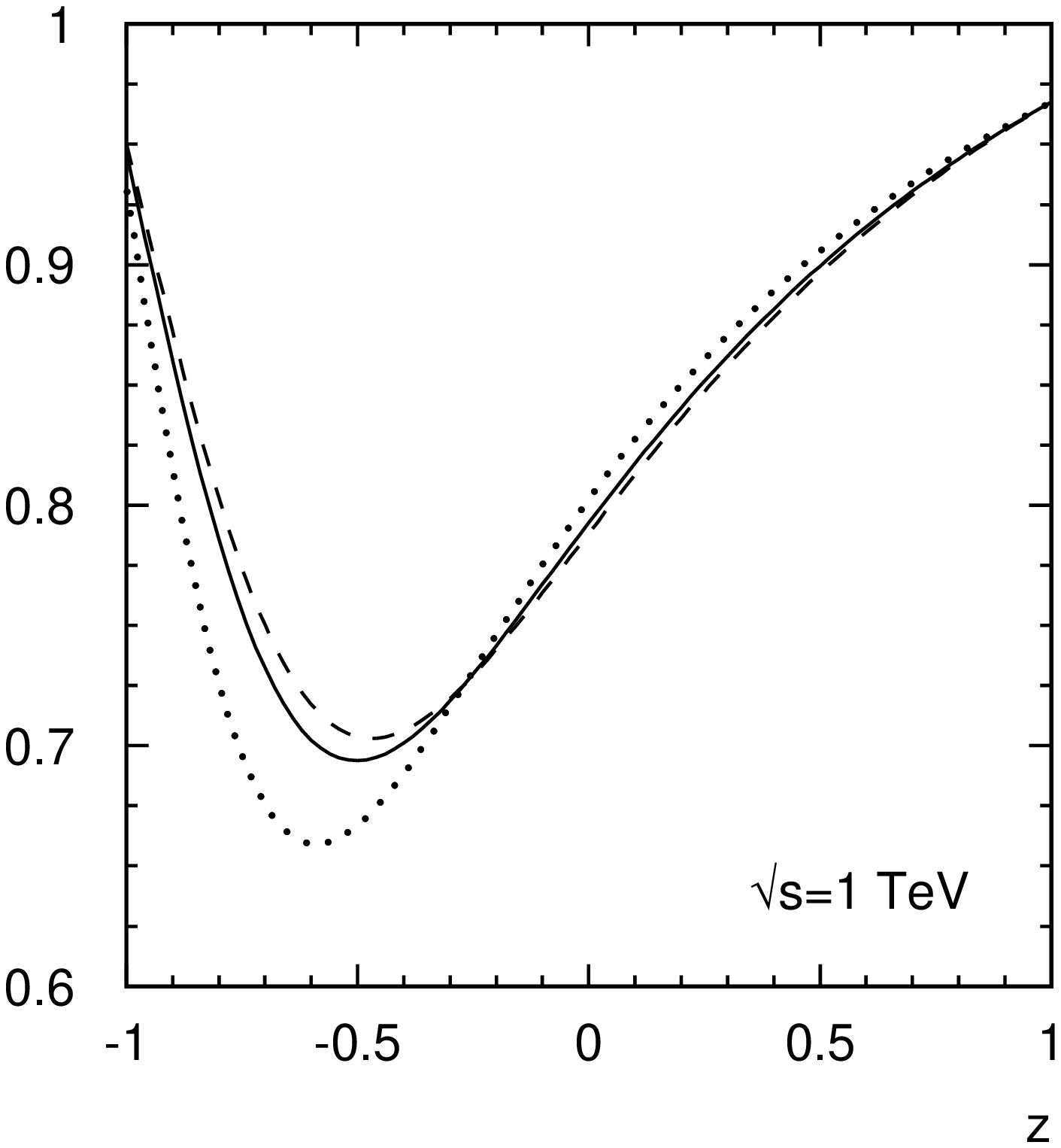}}
\mbox{\epsfysize=41mm\epsffile{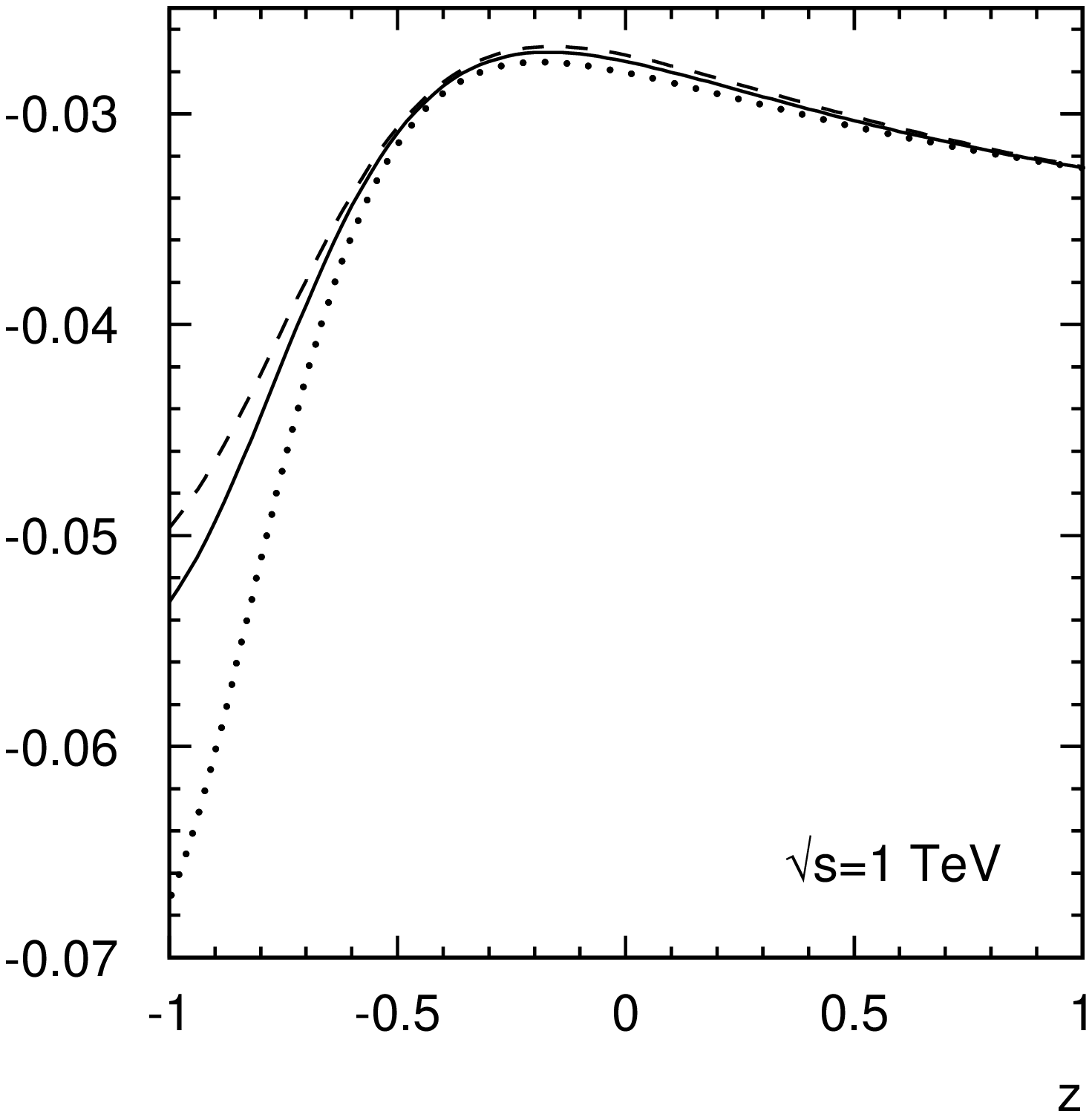}}
\caption{Same as Fig. 1 , but for the correlation
$\hat{k}_iC_{ij}\hat{k}_j$.}
\label{F6}
\end{figure}
We now turn to the discussion of numerical results obtained from
the exact calculation including the QCD corrections.
We consider unpolarized positron beams and
the three cases $\lambda_-=0,\pm 1$. 
For c.m. energies not too far from the $t\bar{t}$ threshold,
the QCD corrections are quite small. (As mentioned before, the parton model
results presented here can not be used in the threshold region itself,
where the expansion in $\alpha_s$ 
does not make sense.) For example, at $\sqrt{s}=0.4$ TeV 
the corrections are smaller than $0.5\%$ ($1\%$) for the top quark polarization
projected onto the electron beam (top quark direction of flight) for all
scattering angles. Far above threshold, the QCD correction to the polarization
can reach values above $5\%$ for special values of 
$z$ (see right plots of Figs. 1 and 2). The normal
polarization shown in Fig. 3 reaches values of a few percent for not too
high c.m. energies.
\par
Since the production of the top quarks proceeds through a single spin-one
gauge boson, the correlation  $C_{ij}\delta_{ij}=4\langle {\bf S}_t\cdot
{\bf S}_{\bar{t}} \rangle$
is exactly equal to 1 at the Born level, independent of the scattering
angle. Only hard gluon emission leads to a deviation from this result.
The QCD correction to this correlation is therefore extremely small
at $\sqrt{s}=0.4$ TeV due to the phase space suppression (see Fig. 4, left). 
However, at $\sqrt{s}=1$ TeV, the hard gluon emission leads to a substantial 
decrease of this correlation, which exceeds $10\%$ for top quarks emitted 
in the backward direction in the case of right-handed electron beams (Fig. 4, right).   
Fig. 5 shows the `beamline' spin correlation $\hat{p}_iC_{ij}\hat{p}_j$. The QCD corrections to this quantity
are smaller than $1\%$ at $\sqrt{s}=0.4$ TeV and of the order of $5\%$ at 
$\sqrt{s}=1$ TeV (right plot of Fig. 5). Finally, 
Fig. 6 depicts our results for the
correlation
$\hat{k}_iC_{ij}\hat{k}_j$. Note that this correlation 
is at Born level equal to $(-1)$ times the 
`helicity' correlation 
$P_{\ell\ell}=\hat{k}_{t,i}C_{ij}\hat{k}_{\bar{t},j}$. This 
special spin correlation, averaged
over the scattering angle, was computed analytically to order $\alpha_s$ 
in ref. \cite{TuBePe98,GrKoLe98}. Further results for other 
c.m. energies and for additional spin observables can be found in 
ref. \cite{BrFlUw99}. 
\section{Conclusions}
At a future linear collider, it will be possible to precisely 
study the rich phenomenology of top quark spin effects, both 
at threshold and in the continuum. Theoretical predictions
for the top quark polarization and the $t\bar{t}$ spin correlations
above threshold are available to order $\alpha_s$. The QCD corrections
are in general small not too far away from threshold, but can reach, for
energies around 1 TeV, values of the order of 5\% or larger in certain
kinematic regions. Their inclusion is mandatory in searches 
for  nonstandard interactions of the top quark.

\bigskip
{\small We would like to thank W. Bernreuther, M. Je\.{z}abek, J.H. K\"uhn, and
T. Teubner for discussions. A.B. would like to thank the organizers of the
Spin 99 conference for their kind hospitality.}

\end{document}